\documentclass[prl,reprint,nofootinbib,superscriptaddress,preprintnumbers,showpacs]{revtex4-1}
\usepackage{amsfonts}
\usepackage{amssymb}
\usepackage{amsmath}
\usepackage{graphicx,color}
\usepackage{dcolumn}
\usepackage{epsfig}
\usepackage{bm}
\usepackage{bbm}
\usepackage{ulem}
\usepackage{hyperref}
\usepackage{lineno}

\usepackage{color}

\newcommand{\ket}[1]{|{#1}\rangle}
\newcommand{\bra}[1]{\langle{#1}|}
\newcommand{\inp}[2]{\langle{#1}|{#2}\rangle}

\def\gtap{\ \raise.3ex\hbox{$>$\kern-.75em\lower1ex\hbox{$\sim$}}\ }
\def\ltap{\ \raise.3ex\hbox{$<$\kern-.75em\lower1ex\hbox{$\sim$}}\ }
\begin{document}

\title{
Impact of final state interactions on
neutrino-nucleon pion production cross sections
extracted from neutrino-deuteron reaction data
}
\author{S. X. Nakamura}
\affiliation{
University of Science and Technology of China, Hefei 230026, 
People's Republic of China
}
\affiliation{
Laborat\'orio de F\'isica Te\'orica e Computacional - LFTC, 
Universidade Cruzeiro do Sul, S\~ao Paulo, SP 01506-000, Brazil
}
\author{H. Kamano}
\affiliation{Research Center for Nuclear Physics, Osaka University, Ibaraki, Osaka
567-0047, Japan}
\author{T. Sato}
\affiliation{Research Center for Nuclear Physics, Osaka University, Ibaraki, Osaka
567-0047, Japan}
\affiliation{J-PARC Branch, KEK Theory Center, IPNS, KEK, Tokai, Ibaraki 319-1106, Japan}

\preprint{ LFTC-18-15/36, J-PARC-TH-0139}

\begin{abstract}
The current and near-future neutrino-oscillation experiments require
significantly improved neutrino-nucleus reaction
models.
Neutrino-nucleon pion production data 
play a crucial role to 
validate corresponding elementary amplitudes that go into 
such neutrino-nucleus models.
Thus the currently available data 
extracted from charged-current neutrino-deuteron reaction data 
($\nu_\mu d\to \mu^-\pi NN$) 
must be corrected for nuclear effects such as the Fermi motion
and final state interactions (FSI).
We study $\nu_\mu d\to \mu^-\pi NN$
with a theoretical model including the impulse mechanism supplemented by
FSI from $NN$ and $\pi N$ rescatterings.
An analysis of the spectator momentum distributions reveals that 
the FSI effects significantly reduce the spectra over the quasifree
peak region, and leads to a useful recipe to 
extract information of
elementary $\nu_\mu N\to \mu^-\pi N$ processes
using $\nu_\mu d\to \mu^-\pi NN$ data,
with the important FSI corrections taken into account.
We provide 
$\nu_\mu N\to \mu^-\pi N$ total cross sections by correcting
the deuterium bubble chamber data for the FSI and 
Fermi motion.
The results 
will bring a significant improvement on
neutrino-nucleus reaction models for
the near-future neutrino-oscillation experiments.

\end{abstract}


\maketitle

The current frontier of neutrino-oscillation experiments,
such as the T2K~\cite{t2k} and the DUNE~\cite{dune},
is to determine the charge-parity violating phase
($\delta_{\rm CP}$) and the neutrino mass hierarchy
as the primary objective.
In this era of precision neutrino experiments, 
we must improve the current situation that
the uncertainty in our knowledge of
 neutrino-nucleus cross sections
in the few-GeV neutrino energy region
is one of the largest sources of systematic uncertainty 
in extracting the oscillation parameters from the data~\cite{white,mahn}.

\begin{figure*}[t]
\begin{center}
\includegraphics[width=0.8\textwidth]{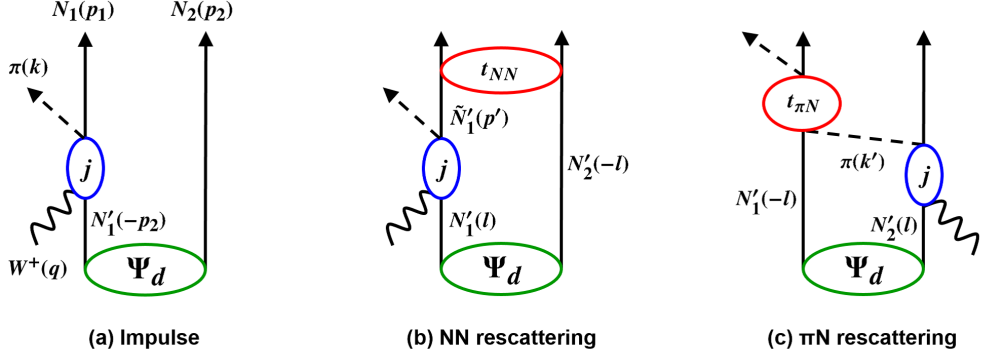}
\end{center}
\caption{
Mechanisms in our $\nu_\mu d\to \mu^-\pi NN$ model:
 (a) impulse; (b) $NN$ rescattering; 
 (c) $\pi N$ rescattering mechanisms.
 Particle labels and their momenta (in parentheses) are defined.
}
\label{fig:deuteron}
\end{figure*}

A reliable model for the elementary neutrino-nucleon reactions is
a key ingredient in 
neutrino-nucleus interaction generators such as 
NEUT~\cite{neut}, GENIE~\cite{genie}, NuWro~\cite{nuwro}, and
GiBUU~\cite{gibuu}, to be used in
neutrino-oscillation analyses.
Many microscopic models~\cite{microscopic} for the neutrino-nucleon single pion
productions ($\nu N\to l\pi N$) have been developed 
with different dynamical contents~\cite{nu_model_pi,SUL,dcc_nu};
see Ref.~\cite{nu-review} for an overview of these 
microscopic models,
and Ref.~\cite{pi_comp} for a detailed comparison.
The common procedure in developing all the models is 
to adjust
the dominant $\Delta(1232)$-excitation mechanism
to fit the total cross section data~\cite{anl,bnl}
of $\nu_\mu p\to \mu^-\pi^+ p$,
$\nu_\mu n\to \mu^-\pi^+ n$, and $\nu_\mu n\to \mu^-\pi^0 p$.
These currently available data for $\nu_\mu N\to \mu^-\pi N$ were  
extracted from neutrino-deuteron reaction ($\nu_\mu d\to \mu^-\pi NN$)
data, 
assuming the quasifree mechanism [Fig.~\ref{fig:deuteron}(a)].
In order to reduce the systematic uncertainty of the neutrino-oscillation
parameters,
an urgent task is to clarify the effects of 
the final state interactions (FSI), 
such as the nucleon and pion
rescattering processes [Fig.~\ref{fig:deuteron}(b,c)],
and then correct the extracted $\nu_\mu N\to \mu^-\pi N$ cross sections.

In this paper,
we analyze the spectator momentum distributions
of $\nu_\mu d\to \mu^-\pi NN$, 
the data of which would be obtainable 
from a possible future neutrino-deuteron 
experiment~\cite{int_ws}.
Taking account of
 the important FSI corrections, we then
find a useful recipe 
for extracting the information of
$\nu_\mu N\to \mu^-\pi N$ processes from the deuteron-target data.
We will also provide 
improved $\nu_\mu N\to \mu^-\pi N$ total cross sections by
correcting
the bubble chamber data~\cite{reanalysis1,reanalysis2},
which are free from the flux uncertainty, 
for the FSI and Fermi motion effects.
The improved data will enable one to develop a more accurate
$\nu_\mu N\to \mu^-\pi N$ model to be used
in extracting the neutrino properties from the oscillation
experiments of the precision era.

The first attempt towards understanding the FSI effects on
$\nu_\mu d\to \mu^-\pi NN$ differential cross sections
has been made recently~\cite{wsl15},
and sizable FSI effects were found
in the quasifree $\Delta(1232)$-production region.
However, their analysis 
focused on this particular kinematical region,
and thus understanding the FSI corrections
on the $\nu_\mu N\to \mu^-\pi N$ data~\cite{anl,bnl} is beyond their scope.
Our calculation will cover the whole phase space of $\nu_\mu d\to \mu^-\pi NN$
with the Monte Carlo phase-space integral.
Our $\nu_\mu d\to \mu^-\pi NN$ reaction model
consists of the relevant mechanisms depicted in Fig.~\ref{fig:deuteron}:
(a) impulse mechanism;
(b) $NN\to NN$ FSI mechanism;
(c) $\pi N\to\pi N$ FSI mechanism.
We use
elementary weak pion production amplitudes and $\pi N\to\pi N$
scattering amplitudes
generated from a
dynamical coupled-channels (DCC) model~\cite{dcc_nu,nu-review,knls13,knls16}.
The DCC model has been developed for
a unified description of
the hadronic and electroweak
reactions on the single nucleon
in the nucleon resonance region: $\pi N, \gamma^{(*)} N\to X$ and
$\nu_l N\to l^- X$ with $X=\pi N, \pi\pi N, \eta N, K\Lambda, K\Sigma$.
The model has been extensively tested by a large amount of data
($\sim$27,000 data points) on 
$\pi N, \gamma N\to \pi N, \eta N, K\Lambda$, $K\Sigma$~\cite{knls13,knls16},
and also by data on electron-induced reactions~\cite{dcc_nu}
for $W \ltap 2$~GeV ($W$: invariant mass of the hadron system)
and $Q^2 \le$ 3~GeV$^2$.
The DCC elementary amplitudes are particularly suited for describing
the neutrino-deuteron reactions that include loop diagrams of hadronic rescatterings.
 This is because the DCC model possesses unique features, which
the other microscopic models do not, such as:
(i) a consistent description of the weak pion productions and $\pi N$
reactions
satisfying two-body as well as three-body
unitarity requirements;
(ii) off-shell amplitudes are available by construction.
The latter feature is crucial for embedding the elementary amplitudes into the
deuteron reaction model in a manner consistent with the multiple scattering
theory.
The deuteron wave function and $NN$ scattering amplitudes are generated
from a realistic $NN$ potential; here we employ the CD-Bonn potential~\cite{cdbonn}.

We have already studied
$\gamma d\to \pi NN$~\cite{gd-pinn,gd-pinn2} and $\gamma d\to \eta pn$~\cite{gd-epn}
using a similar DCC-based model for
meson photoproductions and have shown that
significant FSI effects bring
model {\it predictions} into a good agreement with the data.
These results validate the predictive power of our approach,
and allow us to estimate the FSI effects on
the neutrino-induced pion productions with a
good level of reliability.

The cross sections for charged-current
neutrino-deuteron reactions in the laboratory frame are given as
\begin{eqnarray}
\frac{d\sigma}{d\Omega_{l'} dE_{l'}} & = &
 \left(\frac{G_F V_{ud}}{\sqrt{2}}\right)^2
\frac{1}{4\pi^2}\frac{|\vec{p}_{l^\prime}|}{ |\vec{p}_l|}
 L^{\mu\nu}W_{\mu\nu}\ , \label{cc-crs}
\end{eqnarray}
where $G_F$ and $V_{ud}$ are
the Fermi coupling constant and CKM matrix element, respectively.
The lepton tensor $L^{\mu\nu}$ is given with 
the initial neutrino ($p_l$) and the final lepton ($p_{l'}$) momenta.
The hadron tensor is defined by
\begin{eqnarray}
W^{\mu\nu}&=&
{\sum_{\bar{i},f}}(2\pi)^3
\delta^{(4)}(p_i+q-p_f)
\langle f|J^\mu|i\rangle \langle f|J^\nu|i \rangle^* \ ,
\label{eq:hadron-t}
\end{eqnarray}
where 
$p_i$ ($p_{f}$) is
the total four-momentum of the initial (final) hadron state
and $q=p_l-p_{l'}$;
the average (sum) over the initial (final) hadron states
is denoted by $\sum_{\bar{i}}$ ($\sum_f$).
The hadron current matrix element, 
$\langle f|J^\mu|i\rangle$, 
includes the impulse,
$NN$ FSI, and 
$\pi N$ FSI 
mechanisms
which are given more explicitly with
the particle labels and momenta defined in Fig.~\ref{fig:deuteron} as 
\begin{figure*}[t]
\begin{center}
\includegraphics[width=1\textwidth]{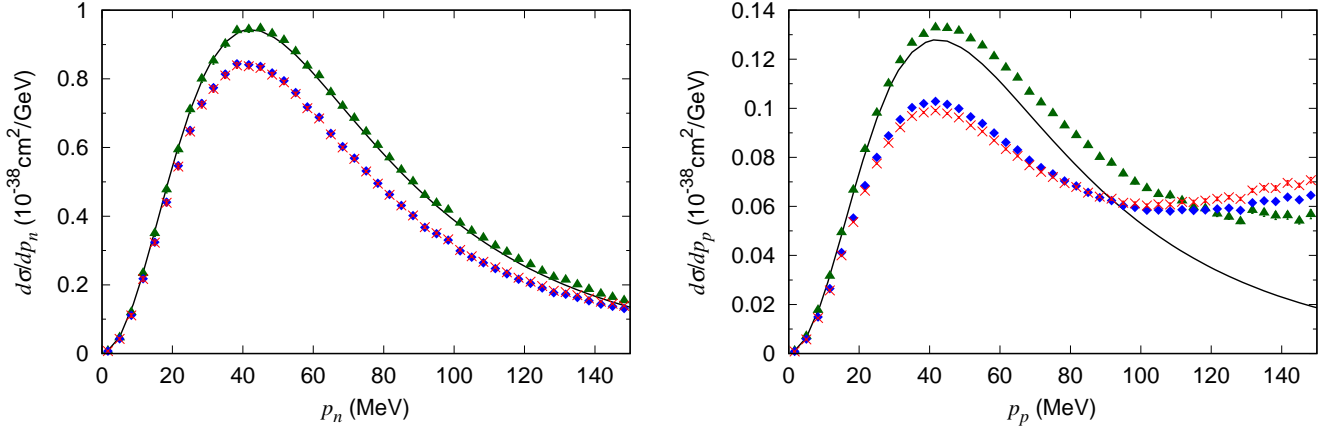}
\end{center}
\caption{
Neutron (left) and proton (right) momentum distributions
in $\nu_\mu d\to\mu^-\pi^+pn$ at $E_\nu=0.5$~GeV.
The green triangles (blue diamonds) [red crosses] are calculated with
 the impulse (impulse + $NN$ rescattering)
[impulse + $NN$ + $\pi N$ rescattering] mechanisms.
The errors are only statistical from the Monte Carlo integral, and are
not shown when smaller than the point size.
The black solid curve in the left [right] panel is
$\nu_\mu p\to\mu^-\pi^+p$ [$\nu_\mu n\to\mu^-\pi^+n$] cross sections
convoluted with the deuteron wave function as in Eq.~(\ref{eq:conv}).
}
\label{fig:p_spct}
\end{figure*}
\begin{widetext}
\begin{eqnarray}
\langle f|J^\mu_{\rm imp}|i\rangle
 &=&
\sqrt{2}
\sum_{s_1',t'_1}
\bra{\pi(\bm{k},t_\pi)\, N_1(\bm{p}_1,s_1,t_1)} 
t_{\pi N,W^+ N}(M_{\pi N_1})
\ket{W^+(\bm{q},\lambda)\, N_1'(-\bm{p}_2 ,s_1',t'_1)}
\nonumber\\
&&\times
\inp{N_1'(-\bm{p}_2,s_1',t'_1)\, N_2 (\bm{p}_2,s_2,t_2)}{\Psi_d(s_d)}
\ ,
\label{eq:amp_imp}
\\
\langle f|J^\mu_{NN}|i\rangle
&=&\sqrt{2}\!\!\!\!
\sum_{s_1',\tilde s_1',s_2',t'_1}
\int d\bm{l}\
\bra{N_1(\bm{p}_1,s_1,t_1)\, N_2(\bm{p}_2,s_2,t_2)}
t_{NN,NN}(M_{N_1N_2}) 
\ket{\tilde N'_1(\bm{q}-\bm{k}+\bm{l},\tilde s'_1,t_1)\, N'_2(-\bm{l},s'_2,t_2)}
\nonumber\\
&&\times
{\bra{\pi(\bm{k},t_\pi)\, \tilde N_1'(\bm{q}-\bm{k}+\bm{l},\tilde s_1',t_1)} 
t_{\pi N,W^+ N}(W)
\ket{W^+(\bm{q},\lambda)\, N_1'(\bm{l},s_1',t'_1)}
\over E-E_N(\bm{q}-\bm{k}+\bm{l})-E_N(-\bm{l})-E_\pi(\bm{k})+i\epsilon}
\nonumber\\
&&\times
\inp{N_1'(\bm{l},s_1',t'_1)\, N_2' (-\bm{l},s_2',t_2)}{\Psi_d(s_d)}
\ ,
\label{eq:amp_NN}
\end{eqnarray}
\end{widetext}
\begin{widetext}
\begin{eqnarray}
\langle f|J^\mu_{\pi N}|i\rangle
 &=& \sqrt{2}
\sum_{s_1',s_2'}
\sum_{t_1',t_2',t'_\pi}
\int d\bm{l}
\bra{\pi(\bm{k},t_\pi)\, N_1(\bm{p}_1,s_1,t_1)}
t_{\pi N,\pi N}(M_{\pi N_1})
\ket{\pi(\bm{q}-\bm{p}_2+\bm{l},t'_\pi)\, N_1'(-\bm{l},s'_1,t_1')}
\nonumber\\
&&\times
{\bra{\pi(\bm{q}-\bm{p}_2+\bm{l},t'_\pi)\, N_2(\bm{p}_2,s_2,t_2)} 
t_{\pi N,W^+ N}(W)
\ket{W^+(\bm{q},\lambda)\, N_2'(\bm{l},s_2',t_2')}
\over 
E-E_N(\bm{p}_2)-E_N(-\bm{l})-E_{\pi}(\bm{q}-\bm{p}_2+\bm{l})+i\epsilon}
\nonumber\\
&&\times
\inp{N_1'(-\bm{l},s_1',t_1')\, N_2' (\bm{l},s_2',t_2')}{\Psi_d(s_d)}
\ ,
\label{eq:amp_MN}
\end{eqnarray}
\end{widetext}
 and the exchange terms that can be obtained 
from Eqs.~(\ref{eq:amp_imp})-(\ref{eq:amp_MN}) 
by flipping the overall sign 
and interchanging 
all subscripts 1 and 2 for the nucleons in the intermediate and final $\pi NN$ states.
Here, the deuteron state with spin projection $s_d$ is denoted as $\ket{\Psi_d(s_d)}$;
$\ket{N(\bm{p},s,t)}$ the nucleon state with momentum $\bm{p}$ and spin and isospin projections 
$s$ and $t$;
$\ket{W^+(\bm{q},\lambda)}$ the $W^+$ boson state with momentum $\bm{q}$ and polarization $\lambda$;
$\ket{\pi(\bm{k},t_\pi)}$ the pion state with momentum $\bm{k}$ and 
the isospin projection $t_\pi$.
The total energy in the laboratory frame is denoted by $E$, and 
the energy of a particle $x$ with the mass $m_{x}$ is
$E_{x}(k)=\sqrt{m_{x}^2+k^2}$.
The two-body invariant masses $M_{\pi N_1}$ and $M_{N_1 N_2}$ are
determined by the final states while 
$W=\sqrt{[E-E_{N}(-\bm{l})]^2-(\bm{l}+\bm{q})^2}$ is used
for the intermediate two-body invariant mass.

The elementary amplitudes $\bra{\pi N} t_{\pi N,W^+ N} \ket{W^+ N'}$ of the weak
pion production~\cite{dcc_nu} and $\bra{\pi N}t_{\pi N,\pi N}\ket{\pi'N'}$ of the
pion-nucleon scattering~\cite{knls13,knls16}
in Eqs.~(\ref{eq:amp_imp})-(\ref{eq:amp_MN}) are first generated from
the DCC model in the two-body CM frame, and then boosted to the
laboratory frame. 
The same frame-transformation procedure is also needed to calculate
the matrix element $\bra{NN}t_{NN,NN}\ket{NN}$ of $NN$ scattering~\cite{cdbonn} in Eq.~(\ref{eq:amp_NN}). 
The  formulas for calculating these matrix elements in 
the laboratory frame are given in Appendix~A in Ref.~~\cite{gd-pinn}.

\begin{figure*}[t]
\begin{center}
\includegraphics[width=1\textwidth]{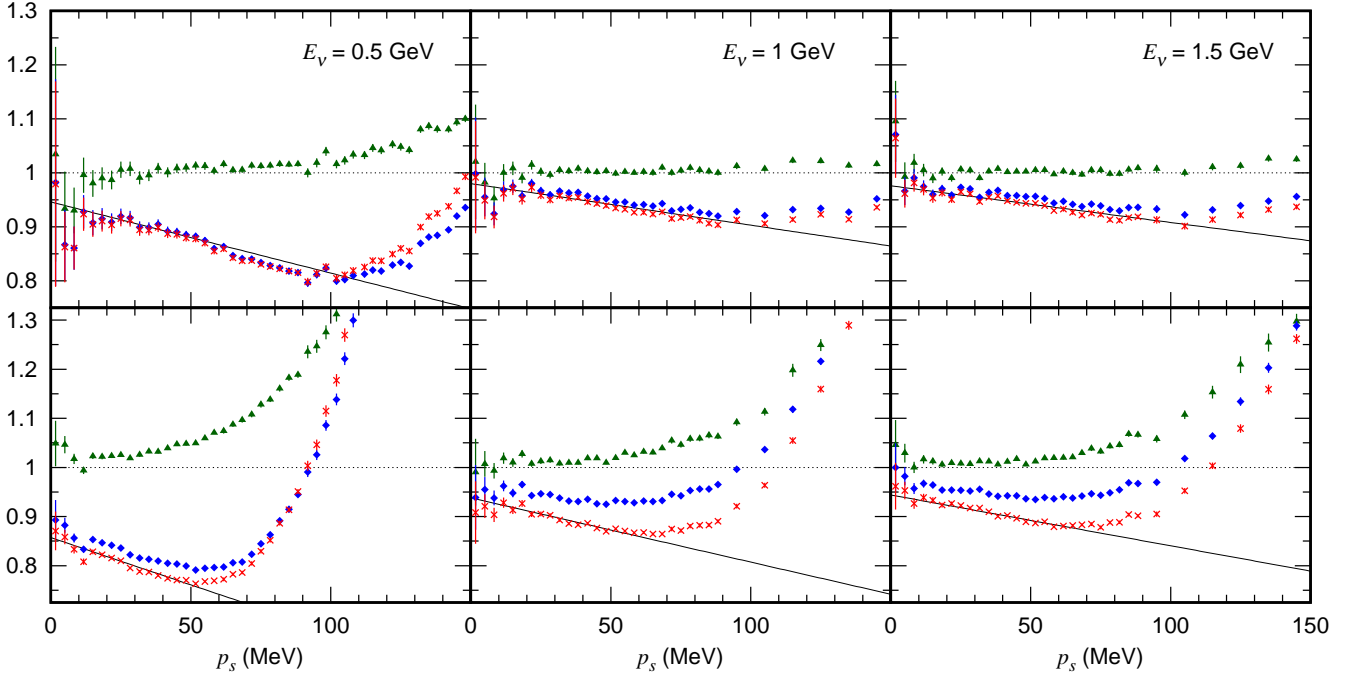}
\end{center}
\caption{
Coefficients $N_\alpha(E_\nu,p_s)$, defined in Eq.~(\ref{eq:nfactor}),
for $\nu_\mu d\to\mu^-\pi^+pn$;
$\alpha=\nu_\mu p\to\mu^-\pi^+p$ ($\nu_\mu n\to\mu^-\pi^+n$)
and $p_s=p_n\, (p_p)$ for the upper (lower) panels.
The neutrino energy is
$E_\nu=0.5$, 1, and  1.5~GeV 
for the left, central, and right panels, respectively.
The analytic functions $N_\alpha^{\rm fit}(E_\nu,p_s)$, 
which are defined in
Eqs.~(\ref{eq:nfactor_fit})-(\ref{eq:nfactor_fit2})
and fitted to the red crosses for
 $p_s\le 50$~MeV, are also shown by the black solid lines.
The other features are the same as those in Fig.~\ref{fig:p_spct}.
}
\label{fig:p_spct-ratio}
\end{figure*}

As in the bubble chamber experiments~\cite{anl,bnl}, we look into the
spectator momentum ($p_s$) distribution in
$\nu_\mu d\to\mu^-\pi N'N_s$ ($N_s$: spectator).
The cross section for
$\nu_\mu p\to\mu^-\pi^+ p$ ($\nu_\mu n\to\mu^-\pi^+ n$) 
can be extracted from the $p_n$ ($p_p$) distribution in
$\nu_\mu d\to\mu^-\pi^+ p n$
because, in a low-$p_n$ ($p_p$) region,
the quasifree pion production 
on the proton (neutron) is expected to dominate while 
the neutron (proton) would hardly contribute.
If this expectation is right,
the spectator momentum distribution ($d\sigma_{\nu_\mu d}/dp_s$)
calculated with the impulse approximation (IA) 
should be approximated accurately with
the $\nu_\mu N\to\mu^-\pi N'(\equiv\alpha)$ cross section convoluted 
with the deuteron wave function ($\Psi_d$) as
\begin{eqnarray}
 {d\tilde\sigma_{\alpha}(E_\nu)\over dp_s} = p_s^2 \int d\Omega_{p_s}
  \sigma_{\alpha} (\tilde E_{\nu}) |\Psi_d(\vec p_s)|^2
  \ ,
\label{eq:conv}
\end{eqnarray}
where the total cross section $\sigma_{\alpha}$ is calculated
with the same $\nu_\mu N\to\mu^-\pi N'$ amplitudes implemented in the 
$\nu_\mu d\to\mu^-\pi NN$ model.
The boosted neutrino energy $\tilde E_{\nu}$ is obtained from 
$E_{\nu}$ by the same Lorentz transformation that
boosts the struck nucleon with the momentum
$-\vec{p}_s$ to its rest.

Indeed, when the spectator in $\nu_\mu d\to\mu^-\pi^+ pn$
is the neutron as in Fig.~\ref{fig:p_spct}(left)
where $E_\nu=0.5$~GeV, 
the convoluted $\nu_\mu p\to\mu^-\pi^+ p$
cross section (black solid curves)
agrees almost perfectly with 
$d\sigma^{\rm IA}_{\nu_\mu d}/dp_n$ (green triangles). 
On the other hand, 
when the spectator is the proton as in Fig.~\ref{fig:p_spct}(right), 
the convoluted $\nu_\mu n\to\mu^-\pi^+ n$ cross section
 undershoots 
$d\sigma^{\rm IA}_{\nu_\mu d}/dp_p$
 in the quasifree peak region.
As $p_p$ increases,
the difference between them becomes significantly larger.
This difference is due to  the contribution from the
$\nu_\mu p\to\mu^-\pi^+ p$ mechanism 
which is responsible for $\sim 87$\% of 
the $\nu_\mu d\to\mu^-\pi^+ pn$ total cross section from the IA calculation
at this neutrino energy.

The $NN$ FSI significantly reduces
$d\sigma_{\nu_\mu d}/dp_s$ for
$\nu_\mu d\to\mu^-\pi^+ pn$, 
especially around the quasifree peak in the low-$p_s$ region,
as can be seen in  the differences between 
the blue diamonds and green triangles in Fig.~\ref{fig:p_spct}.
On the other hand, the spectator momentum distribution for 
$\nu_\mu d\to\mu^-\pi^0 pp$
is hardly changed by the FSI and, thus, is in good agreement with
the convoluted $\nu_\mu n\to\mu^-\pi^0 p$ cross section.
The distinct differences between the $\pi^+$ and $\pi^0$ productions
in the FSI effects is due to the fact that 
the deuteron ($pp$) bound state can (cannot) be formed 
in the $\pi^+$ ($\pi^0$) production.

For a more quantitative study of 
the above observations on $d\sigma_{\nu_\mu d}/dp_s$,
we define
a coefficient $N_\alpha(E_\nu,p_s)$ by
\begin{eqnarray}
 {d\sigma_{\nu d}(E_\nu)\over dp_s} = N_\alpha(E_\nu,p_s)
 {d\tilde\sigma_{\alpha}(E_\nu)\over dp_s} 
  \ .
\label{eq:nfactor}
\end{eqnarray}
The predicted $N_\alpha(E_\nu,p_s)$ are shown in Fig.~\ref{fig:p_spct-ratio}.
A deviation from  $N_\alpha(E_\nu,p_s)\simeq 1$ indicates the contributions
from neutrino reactions on the other nucleon and/or FSI.
Within the IA (green triangles),
the quasifree $\nu_\mu p\to\mu^-\pi^+p$ dominates in the low-$p_n$ (upper panel),
while the quasifree $\nu_\mu n\to\mu^-\pi^+n$ dominance 
in the low-$p_p$ (lower panel) 
is prevented by 
the stronger $\nu_\mu p\to\mu^-\pi^+p$ channel
and thus
$N_\alpha(E_\nu,p_s)$ quickly deviates from one.

\begin{figure*}[t]
\begin{center}
\includegraphics[width=1\textwidth]{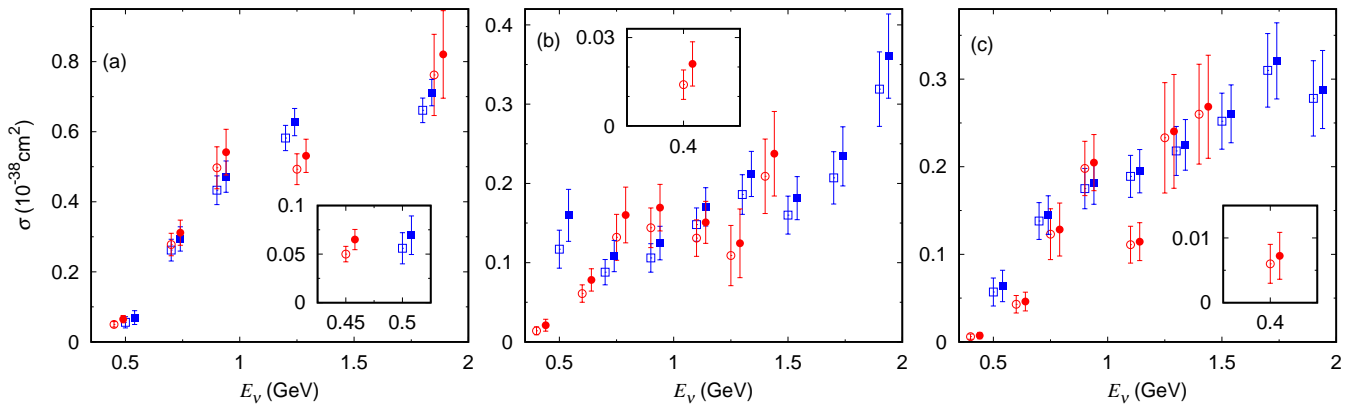}
\end{center}
\caption{
Total cross sections for 
(a) $\nu_\mu p\to \mu^-\pi^+p$, 
(b) $\nu_\mu n\to \mu^-\pi^+n$, and 
(c) $\nu_\mu n\to \mu^-\pi^0p$.
The open red circles (open blue squares) are the reanalyzed ANL (BNL)
data~\cite{reanalysis2} (no $W$ cut),
 which are further corrected for the Fermi motion and FSI and shown by
 the solid red circles (solid blue squares) slightly
[$0.04$~GeV]
 shifted
along the
 $E_\nu$ direction for better visibility. 
The insert shows an enlargement of the small $E_\nu$ region.
}
\label{fig:sig_correct}
\end{figure*}

Figure~\ref{fig:p_spct-ratio} also shows that
the total FSI effects reduce the $p_n$ distribution, 
depending on $p_n$, by 10-20\% (5-10\%) for $E_\nu=0.5$ (1.5)~GeV.
The reduction of the $p_p$ distribution is even larger and,
near the quasifree peak, 
it is $\sim 2$ times larger than the reduction of the $p_n$ spectra.
We also find that the FSI effect depends strongly on the neutrino
energy $E_\nu$.
The $NN$ FSI effects,
which are large at small $NN$ relative energies, 
become smaller as $E_\nu$ increases.
The $\pi N$ FSI effects
become as large as the $NN$ FSI effect for $E_\nu\gtap 1$~GeV
around the quasifree peak of the
$p_p$ spectrum where the $\nu_\mu n\to\mu^-\pi^+ n$  mechanism is
dominant.

\begin{table}[b]
\caption{\label{tab:fit_param}
Parameters for 
$N_\alpha^{\rm fit}(E_\nu,p_s)$ defined in
Eqs.~(\ref{eq:nfactor_fit})-(\ref{eq:nfactor_fit2}).
}
\begin{ruledtabular}
\begin{tabular}{crrr}
$\alpha$& $\nu_\mu p\to\mu^-\pi^+p$&$\nu_\mu n\to\mu^-\pi^+n$&$\nu_\mu n\to\mu^-\pi^0p$\\\hline
$a_{\alpha}$&$  0.9498   $&$ 0.9348  $&$   0.9930  $\\
$b_{\alpha}$&$ -0.5632   $&$-0.7111  $&$  -1.035   $\\
$c_{\alpha}$&$  0.2788   $&$ 0.2211  $&$   0.3142  $\\
$d_{\alpha}$&$ -0.6381   $&$-0.7832  $&$  -1.047   $\\
$e_{\alpha}$&$  0.3171   $&$ 0.2574  $&$   0.3188  $\\
$f_{\alpha}$&$ -0.5496   $&$-0.3170  $&$ -25.18    $\\
$g_{\alpha}$&$  0.0087   $&$-1.387   $&$ -67.88    $\\
$h_{\alpha}$&$ -0.2983   $&$ 0.3030  $&$ 378.5     $\\
\end{tabular}
\end{ruledtabular}
\end{table}
Once $N_\alpha(E_\nu,p_s)$ 
is provided in a simple phenomenological formula,
one can easily extract
$d\tilde\sigma_{\alpha}/dp_s$ of Eq.~(\ref{eq:conv})
from 
$d\sigma_{\nu d}(E_\nu)/dp_s$ data
using Eq.~(\ref{eq:nfactor})
with the FSI effects taken into account.
Thus such a formula
offers a useful recipe: 
one can determine
a $\nu_\mu N\to\mu^-\pi N'$ model by fitting 
the extracted $d\tilde\sigma_{\alpha}/dp_s$.
Also, $N_\alpha(E_\nu,p_s)$ is expected to have a small model dependence
from using our particular DCC-based model, because
it is given by the ratio of  
$d\sigma_{\nu d}/dp_s$ to
$d\tilde\sigma_{\alpha}/dp_s$
both of which are calculated with the same DCC elementary $\nu_\mu N\to\mu^-\pi N'$
amplitudes.
Regarding the functional form,
we find it convenient to use
\begin{eqnarray}
N_\alpha^{\rm fit}(E_\nu,p_s) = A_\alpha(\hat{E}_\nu) + B_\alpha(\hat{E}_\nu) \hat{p}_s \ ,
\label{eq:nfactor_fit}
\end{eqnarray}
with $\hat{E}_\nu\equiv E_\nu/(1\,{\rm GeV})$, 
$\hat{p}_s\equiv p_s/(1\,{\rm GeV})$, and
\begin{eqnarray}
A_\alpha(x) &=&{a_{\alpha}x^2 + b_{\alpha}x + c_{\alpha} \over x^2 +
 d_{\alpha} x + e_{\alpha}  }\ ,\quad
B_\alpha(x) = {f_{\alpha}x + g_{\alpha} \over
 x + h_{\alpha}  } .
\label{eq:nfactor_fit2}
\end{eqnarray}
We fit the parameters ($a_\alpha$--$h_\alpha$) to the numerically
computed $N_\alpha(E_\nu,p_s)$ at several neutrino energies between
$E_\nu=0.4$ and 2~GeV
and in the $p_s\le 50$~MeV region where
the cross sections are dominated by the quasifree process 
aside from the FSI effects.
The obtained parameters are presented 
in Table~\ref{tab:fit_param}.
The fit functions are shown in Fig.~\ref{fig:p_spct-ratio}
by the solid lines
in comparison with the original numerical results.

Our finding on the FSI effects requires a modification 
of the flux-corrected bubble chamber data~\cite{reanalysis1,reanalysis2}.
The data for $d\sigma_{\nu d}/dp_s$ from the $\nu d$ experiments~\cite{anl,bnl}
are unavailable by now.
Therefore, 
we use the $\nu d$ cross section in the quasifree peak region, which
includes a large portion of the total events, 
to estimate the corrections due to the FSI effects.
We integrate Eq.~(\ref{eq:nfactor}) with respect to $p_s$
up to $p_s^{\rm max}=90$~MeV, and then introduce
an effective coefficient $\bar N_\alpha(E_\nu)$
to satisfy the following equation:
\begin{eqnarray}
\bar N_\alpha(E_\nu) \sigma_\alpha(E_\nu)
=
\frac{1}{A}
\int_0^{p_s^{\rm max}}  dp_s {d\sigma_{\nu d}(E_\nu)\over dp_s},
\label{eq:fit}
\end{eqnarray}
with $A \equiv \int_0^{p_s^{\rm max}} d^3p_s |\Psi_d(\vec p_s)|^2$.
The effective coefficient $\bar N_\alpha(E_\nu)$ can be interpreted as 
an average of $N_\alpha(E_\nu,p_s)$, weighted with the nucleon momentum
distribution in the deuteron, over $p_s$.
We may identify $\bar N_\alpha(E_\nu) \sigma_\alpha(E_\nu)$
as the reanalyzed bubble chamber data 
($\sigma^{\rm data}_\alpha(E_\nu)$)~\cite{reanalysis1,reanalysis2},
because both of them can be regarded
as the effective $\nu N$ cross sections extracted from 
$\sigma_{\nu d}(E_\nu)$ without correcting for the FSI and the Fermi
motion.
Because the factor $\bar N_\alpha(E_\nu)$ distorts the true 
$\nu N$ cross sections ($\sigma_\alpha(E_\nu)$) 
by including the FSI and Fermi motion effects,
the corrected data are given by 
$[\bar N_\alpha(E_\nu)]^{-1}\sigma^{\rm data}_\alpha(E_\nu)$.
The corrected data
are shown in Fig.~\ref{fig:sig_correct}~\cite{suppl} in 
comparison with the original ones~\cite{reanalysis2}
to which no $W$ cut has been applied.
The correction is larger for smaller $E_\nu$ 
and enhances the cross sections by scaling factors of
1.05--1.12, 1.10--1.27, and 1.01--1.02 
for 
$\nu_\mu p\to\mu^-\pi^+p$, $\nu_\mu n\to\mu^-\pi^+n$, and
$\nu_\mu n\to\mu^-\pi^0p$, respectively.

In summary,
we have studied for the first time 
$\nu_\mu d\to \mu^-\pi NN$ over the whole phase space
with a reaction model including the FSI mechanisms.
The FSI is found to significantly reduce
the spectator momentum distributions,
depending on the proton or neutron momentum, $\pi^+/\pi^0$ production, and neutrino energy.
We have proposed
a recipe 
to determine an elementary 
$\nu_\mu N\to \mu^-\pi N$ model 
by using data for $d\sigma_{\nu d}(E_\nu)/dp_s$
which could be obtained from
 possible future deuteron-target experiments.
We also presented
the $\nu_\mu N\to \mu^-\pi N$ total cross sections 
by correcting 
the flux-corrected bubble chamber data~\cite{reanalysis2}
for the FSI and Fermi motion.
Because the bubble chamber data
are currently the most important information
for studying the elementary neutrino-induced pion production mechanisms,
the corrected data pave the way to
implementing a significantly improved pion production mechanism 
into a neutrino-nucleus reaction model for
the near-future neutrino-oscillation experiments.
An extension of the present analysis to differential cross sections
such as $W$ and $Q^2$
dependences will be presented elsewhere.

\begin{acknowledgments}
The authors thank T.-S.H. Lee for carefully reading the manuscript and
 giving useful comments.
They also thank C. Wilkinson for providing numerical values of the
 reanalyzed ANL and BNL data.
This work is in part supported by 
National Natural Science Foundation of China (NSFC) under Contract No.~11625523,
by Funda\c{c}\~ao de Amparo \`a Pesquisa do Estado de S\~ao Paulo (FAPESP),
Process No.~2016/15618-8, 
and by JSPS KAKENHI Grant No.~25105010, No.~16K05354, and No.~18K03632.
Numerical computations in this work were carried out
with SR16000 at YITP in Kyoto University,
the High Performance Computing system at RCNP in Osaka University,
the National Energy Research Scientific Computing Center, which is
supported by the Office of Science of the U.S. Department of Energy
under Contract No. DE-AC02-05CH11231, 
and the use of the Bebop and Blues clusters in the Laboratory Computing
Resource Center at Argonne National Laboratory.
\end{acknowledgments}

\appendix

\section{Supplemental material}

The following tables present numerical values
for the total cross sections ($\sigma$) of $\nu_\mu N\to \mu^- N \pi$ 
and their errors ($\delta\sigma$) 
in each neutrino energy bin specified by
the range $[E^{\rm min}_\nu,E^{\rm max}_\nu]$.
The cross sections are obtained
by correcting the reanalyzed ANL and BNL data 
(no $W$ cut)~\cite{reanalysis2}, which are free
 from the flux uncertainty, 
for the final state interactions and the Fermi motion effects.
In each bin, 
the correction factor is calculated at the central value of 
$[E^{\rm min}_\nu,E^{\rm max}_\nu]$.
\begin{table}[h]
\caption{\label{tab:suppl1}
Total cross sections for $\nu_\mu p\to \mu^- p \pi^+$
from correcting the reanalyzed ANL data.
}
\begin{ruledtabular}
\begin{tabular}{cccc}
$E^{\rm min}_\nu$ (GeV)&$E^{\rm max}_\nu$ (GeV)& $\sigma$ ($10^{-38} {\rm cm}^2$)& $\delta\sigma$ ($10^{-38} {\rm cm}^2$)\\
\hline 
   0.3 &  0.6  &   0.065  &   0.010 \\
   0.6 &  0.8  &   0.312  &   0.036 \\
   0.8 &  1.0  &   0.542  &   0.065 \\
   1.0 &  1.5  &   0.531  &   0.047 \\
   1.5 &  2.2  &   0.821  &   0.125 \\
\end{tabular}
\end{ruledtabular}
%
\caption{\label{tab:suppl2}
Total cross sections for $\nu_\mu p\to \mu^- p \pi^+$
from correcting the reanalyzed BNL data.
}
\begin{ruledtabular}
\begin{tabular}{cccc}
$E^{\rm min}_\nu$ (GeV)&$E^{\rm max}_\nu$ (GeV)& $\sigma$ ($10^{-38} {\rm cm}^2$)& $\delta\sigma$ ($10^{-38} {\rm cm}^2$)\\
\hline 
   0.4 &  0.6  &   0.069  &   0.020 \\
   0.6 &  0.8  &   0.294  &   0.035 \\
   0.8 &  1.0  &   0.472  &   0.045 \\
   1.0 &  1.4  &   0.627  &   0.039 \\
   1.4 &  2.2  &   0.712  &   0.038 \\
\end{tabular}
\end{ruledtabular}
\end{table}
\begin{table}[h]
\caption{\label{tab:suppl3}
Total cross sections for $\nu_\mu n\to \mu^- n \pi^+$
from correcting the reanalyzed ANL data.
}
\begin{ruledtabular}
\begin{tabular}{cccc}
$E^{\rm min}_\nu$ (GeV)&$E^{\rm max}_\nu$ (GeV)& $\sigma$ ($10^{-38} {\rm cm}^2$)& $\delta\sigma$ ($10^{-38} {\rm cm}^2$)\\
\hline 
   0.3 &  0.5  &   0.021  &   0.008 \\
   0.5 &  0.7  &   0.078  &   0.014 \\
   0.7 &  0.8  &   0.160  &   0.035 \\
   0.8 &  1.0  &   0.169  &   0.029 \\
   1.0 &  1.2  &   0.151  &   0.026 \\
   1.2 &  1.3  &   0.124  &   0.043 \\
   1.3 &  1.5  &   0.237  &   0.053 \\
\end{tabular}
\end{ruledtabular}
%
\caption{\label{tab:suppl4}
Total cross sections for $\nu_\mu n\to \mu^- n \pi^+$
from correcting the reanalyzed BNL data.
}
\begin{ruledtabular}
\begin{tabular}{cccc}
$E^{\rm min}_\nu$ (GeV)&$E^{\rm max}_\nu$ (GeV)& $\sigma$ ($10^{-38} {\rm cm}^2$)& $\delta\sigma$ ($10^{-38} {\rm cm}^2$)\\
\hline 
   0.4 &  0.6  &   0.160  &   0.033 \\
   0.6 &  0.8  &   0.108  &   0.020 \\
   0.8 &  1.0  &   0.125  &   0.021 \\
   1.0 &  1.2  &   0.170  &   0.024 \\
   1.2 &  1.4  &   0.212  &   0.028 \\
   1.4 &  1.6  &   0.181  &   0.027 \\
   1.6 &  1.8  &   0.234  &   0.037 \\
   1.8 &  2.0  &   0.361  &   0.053 \\
\end{tabular}
\end{ruledtabular}
\end{table}

\begin{table}[h]
\caption{\label{tab:suppl5}
Total cross sections for $\nu_\mu n\to \mu^- p \pi^0$
from correcting the reanalyzed ANL data.
}
\begin{ruledtabular}
\begin{tabular}{cccc}
$E^{\rm min}_\nu$ (GeV)&$E^{\rm max}_\nu$ (GeV)& $\sigma$ ($10^{-38} {\rm cm}^2$)& $\delta\sigma$ ($10^{-38} {\rm cm}^2$)\\
\hline 
   0.3 &  0.5  &   0.007  &   0.004 \\
   0.5 &  0.7  &   0.046  &   0.011 \\
   0.7 &  0.8  &   0.128  &   0.030 \\
   0.8 &  1.0  &   0.205  &   0.032 \\
   1.0 &  1.2  &   0.114  &   0.022 \\
   1.2 &  1.3  &   0.240  &   0.065 \\
   1.3 &  1.5  &   0.268  &   0.059 \\
\end{tabular}
\end{ruledtabular}
%
\caption{\label{tab:suppl6}
Total cross sections for $\nu_\mu n\to \mu^- p \pi^0$
from correcting the reanalyzed BNL data.
}
\begin{ruledtabular}
\begin{tabular}{cccc}
$E^{\rm min}_\nu$ (GeV)&$E^{\rm max}_\nu$ (GeV)& $\sigma$ ($10^{-38} {\rm cm}^2$)& $\delta\sigma$ ($10^{-38} {\rm cm}^2$)\\
\hline 
   0.4 &  0.6  &   0.064  &   0.018 \\
   0.6 &  0.8  &   0.145  &   0.022 \\
   0.8 &  1.0  &   0.181  &   0.024 \\
   1.0 &  1.2  &   0.195  &   0.025 \\
   1.2 &  1.4  &   0.225  &   0.029 \\
   1.4 &  1.6  &   0.260  &   0.033 \\
   1.6 &  1.8  &   0.321  &   0.043 \\
   1.8 &  2.0  &   0.288  &   0.045 \\
\end{tabular}
\end{ruledtabular}
\end{table}


\begin{thebibliography}{}
\bibitem{microscopic}
An alternative approach, 
applicable only at $Q^2\sim 0$,
is to use the PCAC
relation~\cite{dcc_pcac}.
\end{thebibliography}


\end{document}